\documentclass[aps,prb,10pt,twocolumn,amsmath,amssymb,showpacs]{revtex4-1}

\usepackage{graphicx}% Include figure files
\usepackage[usenames]{color} % colors
\usepackage{bm} % bold math
\usepackage{dcolumn} % decimal alignment in tables
\usepackage{soul} % \st{} for strike-out
\usepackage{multirow}
\begin{document}

%===========================%
% TITLE PAGE                %
%===========================%

\title{Topological properties in one-dimensional periodic systems}

\author{Yi-Dong Wu}
\email{wuyidong@ysu.edu.cn}
\affiliation{
Hebei Key Laboratory of Microstructure Material Physics, Yanshan University, Qinhuangdao, Hebei, 066004, China}

\date{\today}

\begin{abstract}
Recently there is trend to study topological properties in one-dimensional(1D) periodic systems. Concepts such as Zak phase are considered as topological invariants that characterize the bulk bands. The bulk 1D systems are classified to topologically nontrivial and trivial phases according to the value of the so-called topological invariant. The existence of edge states or interface states is viewed as a hallmark of the topological nontriviality of the 1D systems. In this work we demonstrate the so-called topological properties in 1D systems are not topological by showing they are description-dependent: the same system can be both topological and trivial depending on how we describe the system. We demonstrate that Zak phase and other related concepts are not topological  invariants by showing they depend on the choice of gauge, especially on the choice of unit cell. We show, for the same bulk system, edge states or interface states can be both present and absent depending the choices of boundaries. So the existence of localized states in 1D system is only a boundary property.
\end{abstract}

\maketitle

\section{Introduction}
Topologically non-trivial phases in periodic systems have attracted great interests recently.\cite{Kane2005Quantum,Kane2005Z2,K2014Quantum,Roth2009Nonlocal}.
Topological edge and surface states have been predicted and observed in two-dimensional(2D) and three-dimensional(3D) systems.
In 2D and 3D systems the topological non-trivialities manifest themselves through the bulk-boundary correspondence. The bulk systems
are characterized by topological invariants such as Chern number in quantum Hall system or $Z_2$ invariants in 2D and 3D topological insulators(TIs).
Bulk topological invariants determine topological property of the gapless edge or surface states.  In a quantum Hall system
 Chern number of bulk bands determines the winding number of edge states\cite{PhysRevLett.71.3697}. In 2D TIs edge states have odd number of pairs of
Fermi points  when $Z_2$ is nontrivial\cite{PhysRevB.76.045302}. In 3D TIs surface states have odd number of Dirac points when $Z_2$ invariant is nontrivial\cite{PhysRevLett.98.106803}.

1D periodic systems have also been used to study the topological properties. Some authors try to use concepts such as Zak phase\cite{PhysRevLett.62.2747} to characterize the topology of an 1D system. They claim that Zak phase can assume two different values $0$ and $\pi$ when the system have certain symmetries. These two values of Zak phase are considered corresponding to two topologically distinct phases of the system. They regard the existence of edge or interface states as a hallmark of topological non-triviality of the 1D system\cite{Xiao,Poli,Cardano,Atala,PhysRevLett.102.065703,PhysRevLett.115.040402,PhysRevB.84.195452,PhysRevLett.110.180403,PhysRevLett.114.123901,PhysRevLett.115.256801,PhysRevX.4.021017,PhysRevLett.108.220401,PhysRevB.91.041402,PhysRevB.92.085118,PhysRevB.93.195420,PhysRevB.94.125119}.  Bulk-edge correspondence in 1D systems is discussed in several works\cite{PhysRevB.84.195452,2017arXiv170506913C,PhysRevX.4.021017}. Zak phase calculated in a specific way is related to the existence or absence of the localized states of the 1D system with specific boundary.

In this paper we demonstrate that Zak phase and other related concepts are not topological invariants by showing they depend on how we describe the system. We also show that the existence of edge or interface states only reflects the choice of boundary and has no topological implication of the bulk system. Some of the previous works are commented to illustrate our point of view.
\section{Zak phase in 1D periodic systems}\label{sec:zakphase}
In this section we discuss the physical meaning of  Zak phase in 1D systems. Both continuous model and tight-binding model are used in the discussion.
\subsection{Continuous model}
In a continuous 1D model Bloch wave function for a given band can be expressed as $\psi_k(x)=e^{ikx}u_k(x)$, where $u_k(x+a)=u_k(x)$ is a periodic function and $a$ is the lattice constant. Zak phase is defined as
\begin{equation}
\varphi_{Zak}=i\int_0^{2\pi/a}\int_{cell}u^*_{k}(x)\frac{\partial}{\partial k}u_{k}(x) dxdk,
\end{equation}
where $u_k(x)$ satisfies $u_{2\pi/a}(x)=\exp(-i2\pi/ax)u_{0}(x)$. The integration over $x$ is on a unit(prime) cell and  $u_{k}(x)$ is normalized by $\int_{cell}u^*_{k}(x)u_{k}(x)dx =1$. Zak phase defined in this way has well physical meaning: it corresponds to the average position of a Wannier function by $\varphi_{Zak}=x_w2\pi/a$, where $x_w$ is the average position of a Wannier function\cite{B2003The}. With this physical interpretation of Zak phase we now show the factors that Zak phase depends on.

First, Zak phase  is gauge-dependent. Under a $U(1)$ gauge transformation $u_k(x)\to\exp(-if(k))u_k(x)$, where $f(k)$ is a continuous real function and $f(2\pi/a)-f(0)=n2\pi$ ($n$ is an integer), $\varphi_{Zak}$ will be transformed to $\varphi_{Zak}+n2\pi$. The new Zak phase corresponding to the average position of a different Wannier function. Since there are infinite number of Wannier functions and the distance between two centers of Wannier functions is always an integral multiple of $a$, Zak phase can only be defined up to an integral multiple of $2\pi$.

Second, Zak phase depends on the origin of coordinate.  If we choose a different origin of coordinate $x$ becomes $x'-d$. Bloch function takes the form of $\psi_k(x')=\exp[ik(x'-d)]u_k(x'-d)=\exp(ikx')\exp(-ikd)u_k(x'-d)$. In the new coordinate $u'_k(x')=\exp(-ikd)u_k(x'-d)$ satisfies $u'_{2\pi/a}(x')=\exp(-i2\pi/ax')u'_{0}(x')$. So Zak phase becomes
\begin{equation}
\begin{split}
\varphi'_{Zak}&=i\int_0^{2\pi/a}\int_{cell}u'^*_{k}(x')\frac{\partial}{\partial k}u'_{k}(x') dx'dk\\&=d2\pi/a+\varphi_{Zak}.
\end{split}
\end{equation}
This result corresponds to the fact that when the origin of coordinate is shifted by $d$ the average position of the Wannier function becomes $x'_w=x_w+d$. In summary, Zak phase (modulo $2\pi$) depends only on the choice of origin of coordinate.

Now we check whether Zak phase is topological or not when the system is inversion symmetric. In Ref. \onlinecite{PhysRevLett.62.2747} it is shown that the centers of the Wannier functions in a 1D inversion symmetric system can only be located at the inversion centers. However, this property is not enough to make Zak phase a topological invariant as claimed in many previous works. If we choose the origin of coordinate arbitrarily, Zak phase can assume any value. From another point of view, in a given coordinate system Zak phase  changes  continuously if we translate the system continously.

If we limit our choice of the origin of coordinate to the inversion centers, there are still two choices in one unit cell. In this case Zak phase can still assume two values $0$ and $\pi$ according to the choices of the origins of coordinates. If two observers try to determine one system is topological or not by measuring the Zak phase of the system, they will reach opposite conclusions just because they choose different inversion centers as origins of their coordinate systems. So Zak phase is not a bulk topological invariant even if the system has inversion symmetry.
\subsection{Tight-binding model}
Tight-binding models are used in most of researches on topological properties of 1D systems. A remarkable difference between tight-binding model and continuous model is that Zak phases calculated with tight-binding models depend on the choices of unit cells. To explain the unit-cell-dependence of Zak phase we explore the physical meaning Zak phase in tight-binding model. We consider a general tight-binding model
\begin{equation}
\hat{H}=\sum\limits_{mnij}t_{ij}(m-n)\hat{C}_{mi}^{\dagger}\hat{C}_{nj},
\end{equation}
where $m$ and $n$ are lattice indexes, $i$ and $j$ denote orbits for a given lattice. $t_{ij}(m-n)=t^*_{ji}(n-m)$ are hopping amplitudes.

We can transform the Hamiltonian to the momentum space by transformation $\hat{C}_{nj}=\int_{-\pi/a}^{\pi/a}\hat{C}_{kj}\phi_k(n)dk$, where $a$ is lattice constant and $\phi_k(n)=\sqrt{\frac{a}{2\pi}}e^{ikna}$ is lattice wave. After the transformation the Hamiltonian becomes
\begin{equation}\label{23}
\hat{H}=\int_0^{2\pi/a}\sum\limits_{ij}\hat{C}_{ki}^{\dagger}T(k)_{ij}\hat{C}_{kj}dk,
\end{equation}
where $T(k)_{ij}=\sum\limits_{n}t_{ij}(n)\exp(-ikn)$. $T(k)_{ij}$ can be considered as elements of a $M\times M$ matrix $T(k)$, where $M$ is the number of orbits in one prime cell. By diagonalizing $T(k)$ we find the eigenenergy spectrum $\epsilon_{\mu}(k)$ and Bloch wave function $|\psi_{\mu k} \rangle=\phi_k(n)|u_{\mu }(k) \rangle$, where $|u_{\mu }(k)\rangle$ is a normalized eigenvector of $T(k)$ and $\mu$ denotes a given band. To construct the Wannier function we require that $|\psi_{\mu k} \rangle$ is periodic with $k$ and thus $|u_{\mu}(k) \rangle$ must be a continuous function of $k$ and satisfies $|u_{\mu}(2\pi) \rangle=|u_{\mu }(0) \rangle$. The normalized Wannier function of the band can be constructed as $|W_{\mu}(n)\rangle= \sqrt{\frac{a}{2\pi}}\int_0^{2\pi/a}\phi_k(n)|u_{\mu }(k) \rangle dk$. For a given $n$, $|W_{\mu}(n)\rangle$ is a $M$ dimensional vector. It can be easily shown that Zak phase
\begin{equation}\label{zak}
\begin{split}
\varphi_{Zak}&=i\int_0^{2\pi/a}\langle u (k)|\frac{\partial}{\partial k}|u (k)\rangle dk\\
             &=\frac{2\pi}{a}\sum\limits_{n}na \langle W_{\mu}(n)|W_{\mu}(n)\rangle=y_w \frac{2\pi}{a}.
\end{split}
\end{equation}
$\langle W_{\mu}(n)|W_{\mu}(n)\rangle$ is the probability the particle in $n$th unit cell. Here we assume orbits with same index $n$ form a unit cell.\\
In calculating $y_w$ all the orbits in $n$th unit cell are assumed to be located at $na$. So $y_w$ is not the actual average position of Wannier function. It is for this reason that Zak phases depends on the choice of unit cell. When we choose a different type of unit cell the same orbit may be considered to be at a different location.

For example, we can make a different choice of unit cell by transformation $\hat{B}_{n'j}=\hat{C}_{n'+\delta_jj}=\hat{C}_{nj}$, where $\delta_j$ are arbitrary integers. If we consider orbits with same $n'$ belong to same unit cell then this choice of unit cell is different from the original one unless all $\delta_j$ are equal. Because there are infinite choices of $\delta_j$ there are infinite choices of unit cell for a given tight-binding model.

It can be demonstrated that the original Wannier function is still a Wannier function with $n'$ as its variable.  So if a particle's wave function is the Wannier function, the probability of the particle in one orbit does not change after the transformation. However, the calculated average position is different because the $nj$ orbit, which is at $na$ originally, is now considered to be located at a $n'a=(n-\delta_j)a$. Thus the Zak phase calculated with the new unit cell is different from the original one.

The new Zak phase are related with the original one by
\begin{equation}\label{ee}
\varphi'_{Zak}=\varphi_{Zak}-2\pi\sum\limits_i\delta_iu_i,
\end{equation}
where $u_i$ is the sum of probability of all $i$th orbits of the Wannier function. $u_i$ is independent of choice of unit cell and choice of gauge. Although it is convenient to group the adjacent orbits into one unit cell, nothing prevents us from using arbitrary unit cells as far as calculating energy band is concerned. By applying Weyl's equidistribution theorem\cite{stein2011fourier} we find that Zak phase can take finitely many values(moduo $2\pi$) when all of $u_i$ are rational numbers. If at least one of $u_i$ is irrational, Zak phase can take infinite number of values(modulo $2\pi$) and these values are equidistributed in $[0,2\pi)$.

It is now clear that the unit-cell-dependence of Zak phase comes from the fact that the positions of the orbits depend on the choice of unit cell. In most systems there is no natural way to choose unit cells. For example the two type unit cells for SSH model in Fig.1 (a) and (b) are equally suitable choices. So we should not attach much physical significance to the Zak phase calculated in this way. Especially it can not be used to study the polarization of electrons as the Berry phase in the continuous model.

\begin{figure}
  \includegraphics[width=8cm]{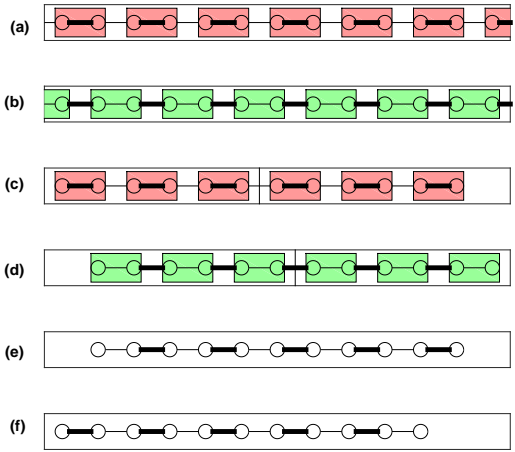}
    \caption{Schematic illustration of SSH model. The circles denote the orbits. Larger hopping amplitudes are represented by thicker lines. (a) and (b) Two choices of unit cells for the same bulk SSH model. Sites in the same box are considered to be in the same unit cell. Red boxes and green boxes correspond to different types of unit cells.  (c)-(f) Four types of boundaries for the same SSH model. Vertical lines in (c) and (d) correspond to inversion centers of the finite systems. In (c) and (d) there are boundary-consistent unit cells. There is no such unit cell in (e) and (f).}
    \label{Fig1}
\end{figure}

If we have the information of the positions of the orbits we can define a Zak phase which has the same physical meaning as Zak phase in the continuous model. The location of the of the orbit can be determined by the relative displacement $d_i$ of the orbit to the lattice. If we positions of the Bravais lattice are fixed the position of $ni$th orbit $na+d_i$ is independent of the choice of unit cells because with a different choice of unit cell $d_i$ also changes so that $n'a+d'_i=na+d_i$.

With given $d_i$ we can expand $\hat{C}_{nj}$ by $\hat{C}_{nj}=\int_{-\pi/a}^{\pi/a}\hat{C}_{kj}\phi_{kj}(n)dk$, where $\phi_{kj}(n)=\sqrt{\frac{a}{2\pi}}e^{ik(na+d_j)}$. That is we use the actual positions of the orbits in the lattice wave $\phi_{kj}(n)$ when expanding $\hat{C}_{nj}$. Then the Hamiltonian becomes $\hat{H}=\int_0^{2\pi/a}\sum\limits_{ij}\hat{C}_{ki}^{\dagger}T'(k)_{ij}\hat{C}_{kj}dk$, where $T'(k)_{ij}=\sum\limits_{n}t_{ij}(n)\exp(-ik(na+d_i-d_j)$. The matrix $T'(k)$ is related to $T(k)$ by a transformation $T'(k)=U(k)T(k)U(k)^{\dag}$, where $U(k)$ is a diagonal matrix with $U_{ii}(k)=\exp(-ikd_i)$. $|u'_{\mu}(k)\rangle=U|u_{\mu}(k)\rangle$ is a normalized eigenvector of $T'(k)$ and we can use it to define Zak phase
\begin{equation}\label{zak}
\phi_{Zak}=i\int_0^{2\pi/a}\langle u'_{\mu} (k)|\frac{\partial}{\partial k}|u'_{\mu} (k)\rangle dk.
\end{equation}
It can be shown $\phi_{Zak}$ correspond to the actual average position of Wannier function. So $\phi_{Zak}$ is independent of the choices of unit cells and it can be used to study the polarization of electrons.

However, even if the positions of the orbits are known, Zak phase is not uniquely determined. When we shift Bravais lattice as a whole, $d_i$ will change and Zak phase changes with $d_i$. It is similar to the dependence of Zak phase on the coordinate origin in the continuous model. There is no natural way to put the Bravais lattice. So only the continuous change of Zak phase is relevant. If the Bravais lattice is fixed, the Zak phase change in a continuous process can be uniquely defined. So $\phi_{Zak}$ can be used to study the polarization of electrons.

In most of tight-binding models we don't have the information of the positions of the orbits, so $d_i=0$ is chosen in most of previous researches. However, this is not the necessary choice when the information of the orbit position is lacking. In fact, we can choose $d_i$ arbitrarily as far as calculating energy bands is concerned and all those choices are equivalent. So choice of $d_i$ can be considered as a general choice of gauge. For example when all $d_i$ are integers it equivalents to choose a new type of unit cell discussed above.

When a general gauge is used, that is not all $d_i$ are integers, $T(k)$ will not be periodic with $k$. It should not come as a surprise because $u_k(x)$ in the continuous model is also not a periodic function of $k$. Various winding numbers of element of $T'(k)$ and eigenvector of $T(k)$ are used to characterize topological property of 1D systems in previous works. However, when $T(k)$ is not periodic with $k$ these winding numbers cannot be defined. So we also should not attach much physical significance to these winding numbers because they are only meaningful when the unphysical assumption that $d_i$ are all integers are made and they depend on the choice of unit cell.
\section{Symmetry and topological property in 1D system}
In this section we will discuss the relationship between symmetry and topological property of 1D systems.  SSH model is used as an example to show that inversion and chiral symmetries do not make the bulk 1D system topological. The $d_i=0$ gauge is used in most of following discussions.

SSH model is widely used to study the topological property of 1D systems. It is claimed the existence of zero-energy edge states is protected by nontrivial Zak phase and chiral symmetry. We consider the tight-binding Hamiltonian for SSH model
\begin{equation}
\hat{H}=\sum\limits_{n}(t_{2n}\hat{C}_{2n+1}^{\dag}\hat{C}_{2n}+t_{2n-1}\hat{C}_{2n}^{\dag}\hat{C}_{2n-1}+\mathrm{H.c.}),
\end{equation}
where $t_{2n}=t$ and $t_{2n-1}=t'$.

There are two possible values of Zak phase of the occupied band $0$ and $\pi$, corresponding to the choices of unit cell in Fig.1 (a) and (b). Although they are considered as two topologically distinct phases in many previous works we can easily see they are just the same bulk system with different choices of unit cells. Now we discuss whether the choices of unit cells can be fixed by considering the symmetry property of the system. If the choice of unit cell can be fixed by symmetries, there may be a symmetry-protected bulk topological property.

We first discuss the space inversion symmetry. In a  bulk system midpoint of each line connecting two adjacent sites can be considered as an inversion center. So there are two inversion centers in each unit cell as the continuous model. Let $\hat{P}_N$ denote the inversion operator whose inversion center is at the midpoint of the line connecting $N$ and $N+1$  sites. Then we have $\hat{P}_N\hat{C}_{N-n'}\hat{P}_N=\eta\hat{C}_{N+n'+1}$, where $\eta$ is a phase factor. It can be shown $\hat{P}_N$ commute with $\hat{H}$.

In a finite system whether there is inversion symmetry or not depends on the boundary and there is only one inversion center if inversion symmetry exists. Fig.1 (c) and (d) show two types  of open boundaries that make the finite systems inversion symmetric.

Here we point out interchanging two orbits in one unit cell is not a symmetry operation of SSH model.  It can be easily demonstrated that the interchange operator  does not commute with $\hat{H}$.

The chiral symmetry defined in Ref.\onlinecite{PhysRevLett.89.077002} can be considered as a consequence of gauge transformation. We rewrite the Hamiltonian of the tight-binding model as
\begin{equation}
\hat{H}=\sum\limits_{ij}(t_{ij}\hat{C}_i^{\dag}\hat{C}_j+\mathrm{H.c.})
\end{equation}
where $i$ and $j$ are integers. Each orbit in the system correspond to an integer $i$. Under a gauge transformation $\hat{C}_i\to e^{ip_i}\hat{C}_i$ the Hamiltonian becomes
\begin{equation}
\hat{H'}=\sum\limits_{ij}(t'_{ij}\hat{C}_i^{\dag}\hat{C}_j+\mathrm{H.c.})
\end{equation}
where $t'_{ij}=t_{ij}e^{i(p_j-p_i)}$.
If there are only nearest-neighbor-hopping between orbits $i$ and $i\pm1$, the phase of $t_{ii-1} $ can be fixed arbitrarily by choosing a suitable gauge.  For example, to add arbitrary phases $e^{i\theta_i}$ to $t_{ii-1}$, we can choose $p_{i-1}-p_i=\theta_i$, that is $p_i=-\sum\limits_{j=1}^i\theta_j+p_0$ when $i>0$ and $p_i=\sum\limits_{j=i+1}^0\theta_j+p_0$ when $i<0$, where $p_0$ is an arbitrary real number. When $\theta_i=\pi$ for all $i$, we have $t'_{ii-1}=-t_{ii-1}$ and thus $\hat{H'}=-\hat{H}$. This property still holds when nonzero hopping between orbits $i$ and $i\pm n$ exists, where $n$ is odd integer. More generally, any Hamiltonian that can be transformed to this form by unitary transformation has chiral symmetry. In this definition of chiral symmetry we do not require the system to be periodic, so it is more general than the chiral symmetry defined in the $k$-space\cite{PhysRevLett.89.077002}.

The SSH model is clearly chiral symmetric. However, we can see from SSH model, chiral symmetry is not a fundamental symmetry as time-reversal symmetry or inversion symmetry and it only reflects the simplification we make in constructing the tight-binding model. For example, if we consider the onsite potential or the next-nearest-neighbour hopping, the chiral symmetry will be broken.\\
From above discussion we can see the symmetries of bulk SSH model are irrelevant to the choice of unit cell. So the symmetry property of SSH model can not determine the choice of unit cell, and thus can not determine Zak phase of the bulk system. Thus we conclude the there are no topologically distinct phases of the bulk SSH model when symmetry property is considered.
\section{Boundary property of 1D systems}
Before discussing 1D systems, we first make some remarks on the bulk-boundary correspondence in 2D and 3D systems. First the topological invariants of the 2D and 3D systems can be defined uniquely without referring to the boundary of the system. The Chern number and the $Z_2$ number are defined on the $k$ space. The wave vector $k$ can only be defined when infinite or periodic boundary condition is used, that is, when the system has no boundary. So they are independent of the boundary of the system. Their definitions are also independent of the choice of unit cells.

Second, the topology of the bulk bands determines the boundary property, instead of the opposite. The boundary property is largely independent of the choice boundary. The only requirement on  the boundary is that it must preserve the symmetry when the topological property is symmetry-protected.

In some of previous works on topological property of 1D systems the situation is just the opposite. They used a boundary-determine-bulk logic: the choice of boundary determines the choice of unit cells which, in turn, determines topology of the bulk system. In fact Schnyder, et al, have pointed out that ``...the quantized values of the Wilson loops do depend on the choice of the unit cell...,the choice of unit cell should be consistent with the location of the boundary..., that the quantized values of Wilson loops in one dimension reflect the boundary physics and are not solely determined from the bulk properties''\cite{PhysRevB.78.195125}.

Here we use the edge state property of SSH model to illustrate the boundary-determine-bulk logic. We have shown that Zak phase of this model can not be determined by considering the symmetry property. So, unlike the 2D and 3D systems, there is no uniquely defined bulk topological invariant in this model. So the so-called topological property of SSH model is meaningless if we do not look at the boundary of the system.

Now we relax the requirement for being a bulk topological property. If symmetry can determine the choice of boundary and one bulk system can only have one type of boundary that preserves the symmetry, then the boundary property may be considered as reflecting the topological property of the bulk. However, this is not the case. The two types of boundaries of the same bulk system shown in Fig.1 (c) and (d) both preserve chiral and inversion symmetry. So there are at least two types of symmetry-preserving boundaries of this system and no one is more suitable to characterize the  topological property of the bulk system.

The boundary-determine-bulk logic begins with a choice of boundary from the two types in Fig.1 (c) and (d). Then they insist that there must be integer number of unit cell in the system, though there is no reason why it is necessary. This requirement will determine the choice of unit cell, which in turn, determines the Zak phase. So it is not the bulk property determines the boundary property, but just the opposite.

The choice of boundary is clearly not a bulk topological property, then why so many authors consider the systems in Fig.1 (c) and (d) as topologically distinct? This is perhaps due to the fact that the two choices of boundaries of the same bulk system can not be continuously connected without breaking the chiral symmetry\cite{PhysRevLett.89.077002}. In this sense the zero-energy edge states are at best a boundary-dependent topological property, that is, they only depend on the choice of the boundary.

The usefulness of this boundary-dependent topological property is very limited. First, to define this property the system must have a boundary. Thus it is meaningless in the pure bulk systems such as an infinite system or a system with periodic boundary condition. However, only in such systems can Zak phase be defined. Second, it can only be defined in a system with integer number of unit cells. For example, it is meaningless in systems shown in Fig.1 (e) and (f) because we do not know the choice of unit cell should be consistent with the left end or right end of the system. One may claim in systems shown in Fig.1(e) and (f) one end is topological and the other end is trivial. However, this conclusion can only confirm the edge states are just boundary property.  Third, it depends strictly on the chiral symmetry. In next section we will show when chiral symmetry is broken, even only at the boundary of the system, the boundary-dependent topological property will become meaningless.

The status of this boundary-dependent property as a topological property can be further diminished. First, we must realize that the orbits outside of the system do not vanish. By using the open boundary condition hopping to these orbits are prohibited\cite{PhysRevLett.89.077002}. In real systems we do not consider these hopping terms not because there are impenetrable barriers at the boundary, simply because the orbits outside of the system have higher energies and thus become unaccessible. If we want construct a more realistic model of the boundary than the open boundary we may assume the orbits outside of the system have large positive onsite potentials. So in real systems chiral symmetries are always broken at the boundaries.

\begin{figure}
  \includegraphics[width=8cm]{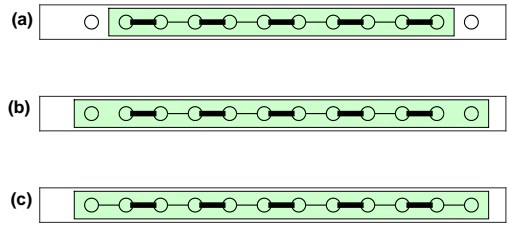}
    \caption{Schematic illustration of boundary property of SSH model. There are two extra isolated sites in (a) and (b). (a) The extra sites are not included in the system. (b) The sites are included in the system. The system in (b) can be continuously connected to (c) without breaking the chiral symmetry.
        }
    \label{Fig1}
\end{figure}

To preserve the chiral symmetry at the boundary we assume the orbits outside of the system also have zero onsite potential and only two extra orbits are considered as shown in Fig.2 (a). If these two orbits are not considered as part of the system(Fig.2 (a)), the system is in the so-called trivial phase. If we include the two sites into the system(Fig.2 (b)), the system will be in the so-called topological states. This can be easily demonstrated by continuously turning on the hopping to the extra sites. Then the system can be continuously connected to the so-called topological phase(Fig.2 (c)) without breaking chiral symmetry. So the same physical reality can be considered both topological and trivial by different observers depending on how they define the system. In this sense the so-called topological property is more subjective than objective.
\section{Some examples}
To illustrate our point of view we comment on some of the previous works. Because there are so many works on this respect we only select those we consider as most typical and influential.
\begin{figure}
  \includegraphics[width=8cm]{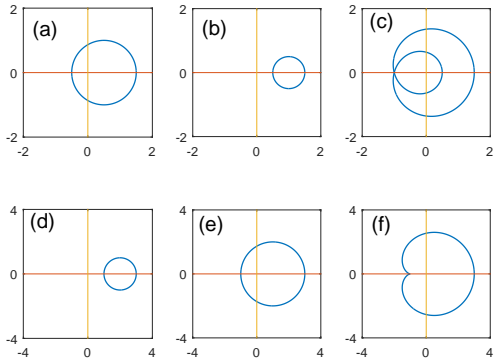}
    \caption{Winding property of $\textbf{g}$(k) in Ref.\onlinecite{PhysRevB.84.195452} with different choices of unit cells. (a)-(c)[(d)-(f)] the so-called topological(trivial) case in Ref. \onlinecite{PhysRevB.84.195452}, where $\delta_1-\delta_2=0,1$ and $-1$ in (a)-(c)[(d)-(f)] respectively. (a) and (d) are reproductions of the results in Ref.\onlinecite{PhysRevB.84.195452}.}
    \label{Fig2}
\end{figure}

Ref.\onlinecite{PhysRevB.84.195452} and Ref.\onlinecite{PhysRevLett.89.077002} are two of the first papers to study the topological property of 1D system with chiral symmetry. As we have shown the so-called topological properties are at best boundary-dependent topological properties. They both use winding numbers in the $k$ space to study the topological property. Mathematically, the winding numbers equivalent to elements of the homotopy group $\pi_1[X]$, where X is a topological space. In this case $X$ is a plane minus the origin.  Like Zak phase the winding numbers depend on the choice of unit cell as shown in Fig.3. Because the unit cell depends on the boundary, the winding number is also a boundary property.

In Ref.\onlinecite{Atala} SSH model is used to study the topological property of 1D cold atom system. In this paper the authors go a step further to claim that there are two phases of SSH model: one is topological and the other is trivial. We find this paper particularly misleading.

First, we explain why their conclusion that Zak phase depends on the choice of unit cell in continuous model is a mistake. The problem of their argument is that they erroneously believe unit cell in continuous model must take the form $x\in[0,a)$, where a is lattice constant. So each time a different unit cell is chosen there must be a change of origin of coordinate so that the starting point of the unit cell is always $0$.\\ However, this is totally unnecessary.  Unit(prime) cell is a volume of space that, when translated through all the vectors in the Bravais lattice, fills all of the space without overlapping itself or leaving voids\cite{ashcroft1976solid}. There are infinite ways of choosing the unit cells, e.g. $x\in[-a/2,a/2)$,$x\in[-a/3,2a/3)$ and $x\in[-a,-a/2)\cup[a/2,a)$. Two choices of prime cells can always be made identical by breaking one unit cell into parts and translating different parts by different Bravais lattices. Because $u_k(x)$ is a periodic function of the Bravais lattice, when a volume is translated by a Bravais lattice the integration of $u^*_{k}(x)\frac{\partial}{\partial k}u_{k}(x)$ on that volume will not change. Thus the integration of  $u^*_{k}(x)\frac{\partial}{\partial k}u_{k}(x)$ is independent of the choice of the unit cell. So Zak phase is independent of the choice of unit cell for a given origin of coordinate.

As shown above, Zak phase only depends on the choice of origin of coordinate. Clearly it is not the choice of the unit cell but the accompanying change of origin of coordinate that changes the Zak phase of the continuous model in Ref.\onlinecite{Atala}. If we change the choice of unit cell without changing the origin of coordinate Zak phase in the continuous model will not change.

\begin{figure}
  \includegraphics[width=8cm]{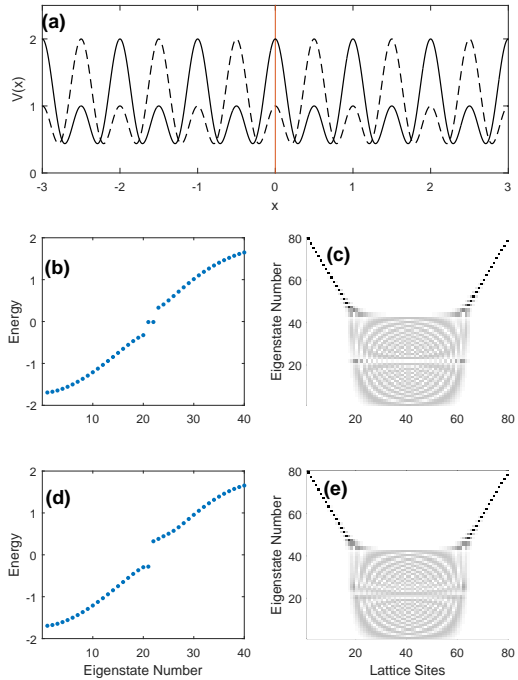}
    \caption{(a)Lattice potential used in Ref.\onlinecite{Atala}. The solid(dashed) line corresponds to $\phi=\pi$($\phi=0$).(b)-(e) Edge state properties of the system in a power law potential. The only difference between our results and those in Ref.\onlinecite{Atala} is $r^0=21$ is used in our work and $r^0=20$ is used in Ref.\onlinecite{Atala}.  (b) and (c) correspond to the so-called trivial phases and (d) and (e) correspond to the so-called topological phase. We see the edge state property is just the opposite to those in the supplementary information of Ref.\onlinecite{Atala}.}
    \label{Fig4}
\end{figure}

In Ref.\onlinecite{Atala} the Zak phase difference between the two configurations is considered as a topological invariant. They also claim ``There is,
however, a natural choice of the origin of the unit cell with which one can identify which dimerization configuration is topologically trivial or non-trivial''.

We will show the Zak phase difference is not a bulk topological invariant by using both the continuous model and tight-binding model. To compare the Zak phases of two systems we must find a way to specify the Zak phase of each system. In the continuous model we must specify the origin of the coordinate of each system. If the origin of the coordinates of the two systems arbitrarily the Zak phase difference can take any value. So in this case it is natural to put the two systems in one coordinate system. However, even if we put them in one coordinate system, there is still no way to get an unique Zak phase difference because the Zak phase difference depends on the relative displacement of the two systems. So the Zak phase difference in the continuous model only reflects how the two systems are put in one coordinate system and has no topological implication at all.

For example, if two identical systems are put in one coordinate system with a relative displacement of half period, the Zak phase difference between the two systems is $\pi$. In fact this is just the Zak phase difference they measured in Ref.\onlinecite{Atala}. They claim the $\phi=0$ and $\phi=\pi$  cases for lattice potential $V(x)=V_1\sin^2(k_1x+\phi/2)+V_s\sin^2(2k_1x+\pi/2)$ are topologically distinct and the occupied bands of two cases have a relative Zak phase $\pi$. However it can be easily seen from Fig.4 (a) that the two cases are just the same lattice potential with a relative translation of half period. Because the bulk systems are identical the Zak phase difference $\pi$ has no topological implication. A trivial system will not become a topological system simply because we shift the system by half period. 

It is more difficult to compare Zak phases of two tight-binding models because the Zak phase of tight-binding model depends on the choices of unit cell and general gauge. In most of previous works the $d_i=0$ gauge is used and orbits in one unit cell are assumed be located at the $na$. In this case Zak phase depends on the choice depends on unit cell. As shown in Fig.1 (a) and (b) two identical bulk SSH systems have a Zak phase difference $\pi$ only because different unit cells are chosen. These two choices of unit cells are equally suitable and no one is more ``natural'' than the other.  So if we only look at the bulk system the Zak phase difference is meaningless.

The choice of the unit cell can only be determined when the boundary-determine-bulk logic is used. Then the two systems in Fig.1 (c) and (d) have a Zak phase difference $\pi$. These two systems are the same bulk system with different choice of boundaries. So if we use the boundary-determine-bulk logic the Zak phase difference only reflects different choices of  boundaries of the same bulk system.  We will show the measured Zak phase in Ref.\onlinecite{Atala} has nothing to do with the boundary property.

In Ref.\onlinecite{Atala} the reason why Zak phase of the tight-binding model is different from other works is not because of the choice of the unit cell as claimed by the authors. They also consider the two orbits with same $n$ form one unit cell. The difference is due to they use a different general gauge discussed above. They assume one of the two orbits in one unit cell to be located at $na$ and the other at $(n+1/2)a$, which corresponds to $d_1=0$ and $d_2=a/2$. This choice of $d_1$ and $d_2$ certainly does not corresponds to the actual position of orbits because the orbits are not equally-spaced in the optical superlattice show in Fig.4 (a). We can not see why such a gauge are chosen. By using this gauge the Zak phase difference between the two choice of unit cell in Fig.1 (a) and (b) is still $\pi$.

To justify the Zak phase difference is topological invariant they also use the boundary-determine-bulk logic. They use the existence and absence of edge states to differentiate the topological phase and   trivial phase. The two choices of boundaries of the same system in Fig.1 (c) and (d) are once again considered topologically distinct.

To confirm the topological distinction of the two systems they claim the absence or presence of edge states can still be observed when when a power law potential is present. Contrary to their conclusion, their discussion only provides an excellent example that edge state property can be continuously changed when chiral symmetry is broken. When there are on-site potential the chiral symmetry is broken as we shown above. In this case the chiral symmetry is significantly broken only at the boundary of the system since the power law potential is almost zero in the bulk. When we continuously vary $r_0$ from $20$ to $21$ we find the edge state property becomes opposite to those shown in the supplementary information of Ref.\onlinecite{Atala}. The edge state property when $r_0=21$ is shown in Fig.4 (b)-(e). In this example we can see the boundary with edge states can be continuously connected to the boundary without edge states when chiral symmetry is broken. Thus there is no topological distinction at all between the two so-called topological phases without the protection of chiral symmetry.  So, instead of justifying the topological distinction of the two phases, this example only show that the boundary-dependent topological property becomes meaningless when the chiral symmetry is broken at the boundary.

As shown above the Zak phase difference can at best reflect the boundary-dependent topological property discussed above. So to observe such a property the boundary of the system must be controlled. 
However, they show no control on the boundary of the system in their experiment. In fact, to observe the boundary-dependent topological property only control the boundary is not enough, we must ensure chiral symmetry is not broken. For example the next-nearest-neighbour hopping must be strictly zero. Nothing about the chiral symmetry appears in their work. So the measured Zak phase difference has nothing to do with the boundary-dependent topological property. It only reflects the fact that the Zak phase change is $\pi$ when the same continuous system is translated by half period in a fixed coordinate system, which can observed in any periodic system. So the measured Zak phase difference has no topological implication at all.

In Ref. \onlinecite{PhysRevB.94.125119} $Z=\phi_{Zak}/\pi$ is considered as a topological invariant and $Z=0,1,2$ corresponds to distinct topological phases. They claim that ``Since the value of Zak phase is gauge dependent, we follow the choice of unit cell so that the Zak phase of Bloch bands takes the values $0$, $\pi$, or $2\pi$.'' The $Z=1(2)$ case is considered to be topologically non-trivial because there are one pair(two pairs) of zero-energy edge states. The $Z=0$ case is consider as topologically trivial because there is no edge state.

\begin{figure}
  \includegraphics[width=8cm]{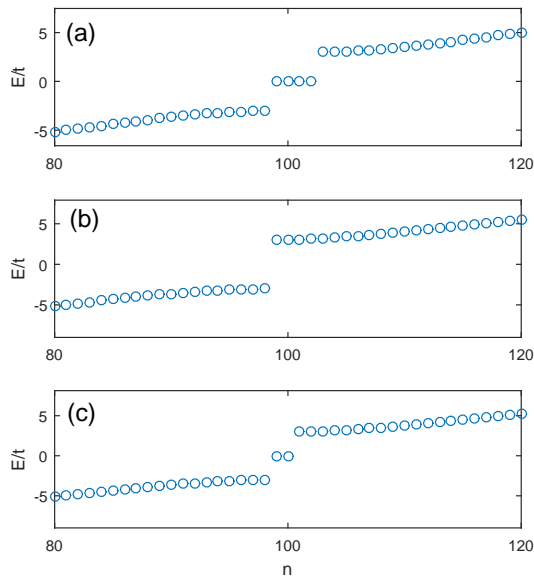}
    \caption{(a)A reproduction of result in Ref. \onlinecite{PhysRevB.94.125119}. The spectrum of the model in Ref. \onlinecite{PhysRevB.94.125119} with open boundary condition, where $\delta t=-0.5t$,$\lambda=3t$ and $\lambda'=9t$ corresponding to the so-called topological case with $Z=2$. (b) and (c) The spectrum of the same model with different boundaries. (b) Spectrum of same model in (a) with the first and last sites deleted.  (c) Spectrum of same model in (a) with the last site deleted.}
    \label{Fig5}
\end{figure}

They seem to be unaware of the $U(1)$ gauge transformation so that even $\phi_{Zak}=0$ and $\phi_{Zak}=2\pi$ are considered as topologically distinct. This is clearly not the case.  There are two occupied bands in their model. Zak phase they use is the sum of the Zak phases of the two bands. We have shown Zak phase of one band can only be defined modulo $2\pi$. So the sum of the Zak phases is also defined modulo $2\pi$. For example, in the $Z=0$ case, we can multiply the wave function of one of the two bands by a phase factor $\exp(-ik)$. Then Zak phase of the occupied bands becomes $2\pi$. So the $Z=0$ and $Z=2$ cases have same Zak phase, and thus they can not be considered as topologically distinct. The existence of the edge state of this model depends on the choice of boundary as is shown in Fig.5. So the so-called bulk topological property is also a boundary property like that in SSH model.

In Ref. \onlinecite{PhysRevLett.110.180403} and Ref.\onlinecite{PhysRevB.91.041402} the so-called bulk topological property depends on how the sites in the systems are labeled. In Ref. \onlinecite{PhysRevLett.110.180403} off-diagonal AAH model is used to study topological property of 1D system. When $b=1/2$ this off-diagonal AAH model
is nothing but a SSH model in disguise: if we let $J=t[1-\lambda\cos(\varphi_{\lambda})]$, $J'=t[1+\lambda\cos(\varphi_{\lambda})]$, $\hat{a}_n=\hat{c}_{2n}$ and $\hat{b}_n=\hat{c}_{2n-1}$, the Hamiltonian will become a SSH type $\hat{H}=\sum\limits_{n}(J\hat{a}_n^{\dagger}\hat{b} _n+J'\hat{b}_{n+1}^{\dagger}\hat{a} _n+h.c.)$. $J>J'$when $-\pi/2<\varphi_{\lambda}<\pi/2$ and  $J'>J$ when $\pi/2<\varphi_{\lambda}<3\pi/2$.

If we relabel the site $n$ by $n+1$, the bulk Hamiltonian of  will become $\hat{H}=\sum\limits_{n}(t[1+\lambda\cos(n\pi+\varphi_{\lambda}-\pi)]\hat{c}_{n+1}^{\dagger}\hat{c} _n+h.c.)$. That is  $\varphi_{\lambda}$ and $\varphi_{\lambda}-\pi$ correspond to the same bulk system with different labeling of the sites. In Ref. \onlinecite{PhysRevLett.110.180403} these two cases are considered as topologically distinct. Now we see there are no topological distinction at all if we only look at the bulk systems. As a consequence, the so-called topological invariant is also not a bulk property. Instead of Zak phase they try to use $Z_2$ index of  Majorana chain to characterize this system. However, this topological invariant is only relevant in the presence of superconductivity\cite{kitaev2001unpaired}. It can be easily show that the same system has different $Z_2$ indexs before and after the relabeling.

In Ref. \onlinecite{PhysRevLett.110.180403} the so-called topological case has $Z_2=1$ and there are zero-energy states. After the relabelling $Z_2=0$ and the system must be in a so-called trivial phase. However, the zero-energy edge states are still there. If the existence of edge states is the only hallmark of the topological non-triviality the system must be in a topological phase. Then why the finite system is topological but $Z_2$ indicates it should be trivial? The answer is that $Z_2$ is not a bulk topological invariant and the existence of edge states only reflect the choice of edge sites.

 \begin{figure}
  \includegraphics[width=8cm]{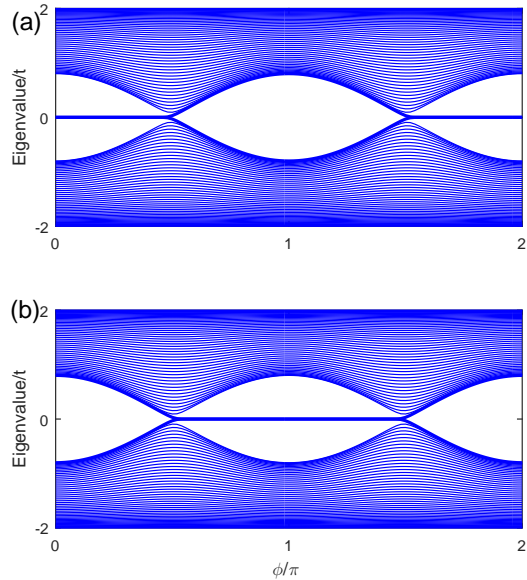}
    \caption{Spectrums of off-diagonal AAH model with $b=1/2$ in Ref.\onlinecite{PhysRevLett.110.180403}. (a)A reproduction of the result depicted in Fig.1 (a) in Ref.\onlinecite{PhysRevLett.110.180403}. (b)Spectrum of the same bulk system with different choice of boundary. The only difference in (a) and (b) is that sites from $n=1$ to $n=100$ are used in (a) and sites from $n=0$ to $n=99$ are used in (b).}
    \label{Fig6}
\end{figure}

As we have explained if we do not look at the boundary of the system $Z_2$ is meaningless because it can assume both $0$ and $1$ depending how the sites are labelled. After the relabelling the edge state property will not change. But the labelling of the edge sites changes. If they are originally labelled as $n=1$ and $n=N$, it is now labelled as $n=2$ and $n=N+1$. That is the edge state property depends on the choice of edge sites. When $n=0$ and $n=N-1$ are chosen as edge sites the $Z_2=0$ case will have zero-energy edge states and the $Z_2=1$ case has no edge state. So $Z_2=0$ will be considered as topological and $Z_2=1$ will be considered as trivial by the author of Ref. \onlinecite{PhysRevLett.110.180403}. There is no reason why $n=1$ and $n=N$ must be chosen as edge sites when we study the topological property of the system. So $Z_2$ is not a topological invariant and the edge state property only reflects the choice of boundary as we have discussed in the SSH model.

As a comparison, the edge state property in Ref. \onlinecite{kitaev2001unpaired} is truly topological. The invariant in Ref. \onlinecite{kitaev2001unpaired} is independent how the sites are labelled and it can be uniquely defined when periodic boundary condition is used. So it is a truly bulk invariant because it can be defined without considering the choice of boundary. The edge state property is also independent of the choice of edge sites.

\begin{figure}
  \includegraphics[width=8cm]{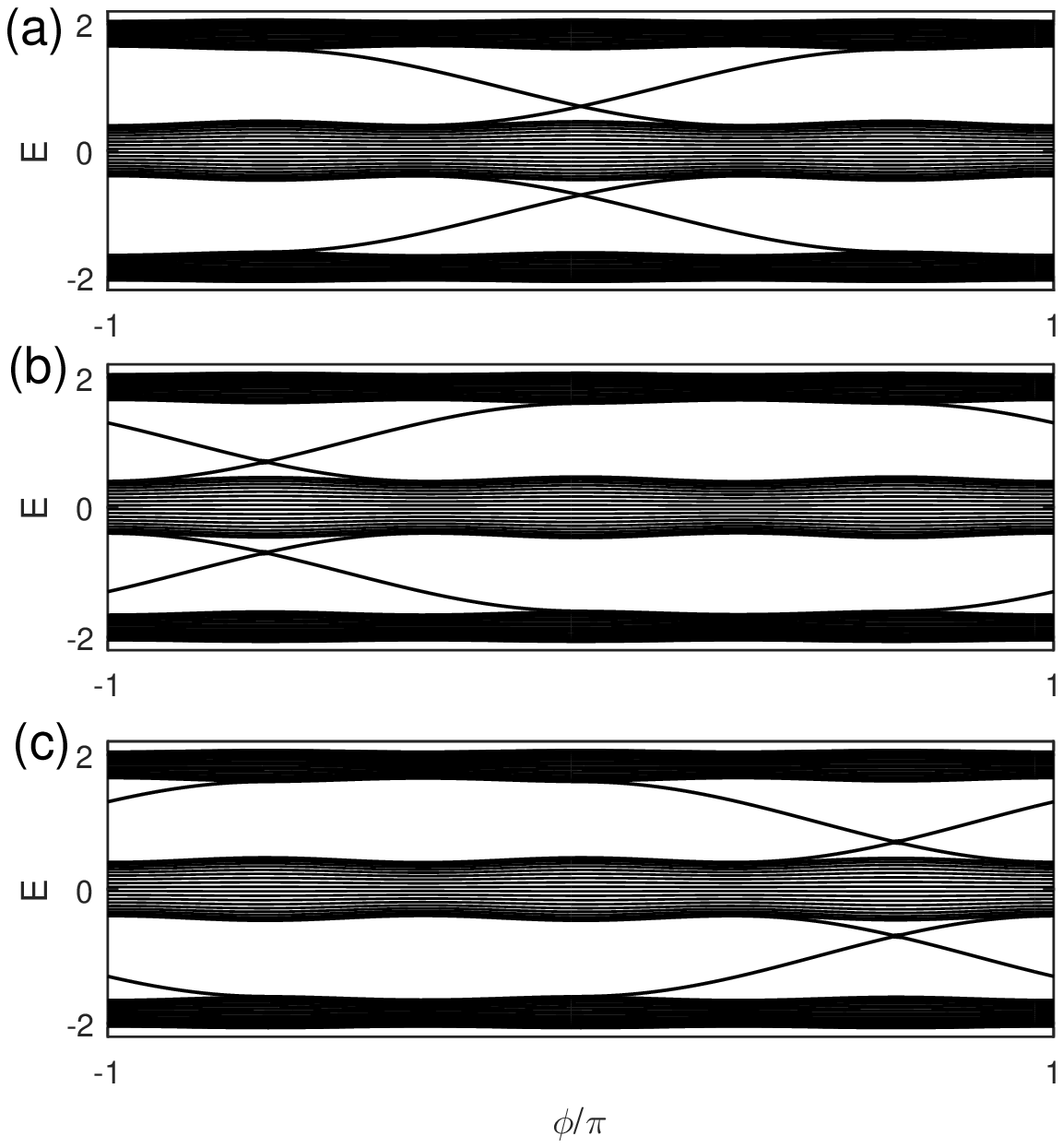}
    \caption{Spectrums of the model in Ref. \onlinecite{PhysRevB.91.041402} with different open boundary conditions, where $T=3$, $\lambda=0.6$. In (a), (b) and (c) sites $j=1\to60$,$j=2\to61$ and $j=0\to 59$ are used respectively. (a) is a reproduction of result depicted in Fig.1 (e) in Ref. \onlinecite{PhysRevB.91.041402}. }
    \label{Fig7}
\end{figure}

In Ref.\onlinecite{PhysRevB.91.041402} the so-called topological property is also label-dependent. In this paper they use periodically modulated hopping amplitudes to study topological property in 1D system. This 1D system is described by the following Hamiltonian
  \begin{equation}
 \hat{H}_1=\sum\limits_{j}(t_{j,j+1}\hat{c}_j^\dag\hat{c}_{j+1}+H.c.)
 \end{equation}
 where $t_{j,j+1}=t(1+\lambda\cos(2\pi j/T+\varphi))$. They use Berry(Zak) phase $\gamma$ to characterize the bulk system. For example $\varphi=0$ in the $T=3$ case is considered as topologically non-trivial. Similar to Ref.\onlinecite{PhysRevLett.110.180403}, $\varphi$ and $\varphi-n2\pi/T$ correspond to the same bulk system, where $n$ is an integer. This can be seen by relabeling the lattice $j$ by $j+n$, then the hopping amplitude will become $t_{j,j+1}=t(1+\lambda\cos(2\pi j/T+\varphi-n2\pi/T)$.  Zak phase of these system depends on how we group the sites into unit cells. Again the choice of unit cells can only by fixed by the choice of boundaries. So the so-called topological property is a boundary property. Thus, it is not surprise to see the edge state property depends on the choice of boundary as is shown in Fig.7.

 Interface states between two 1D systems are also used to study the topological property of 1D systems\cite{PhysRevX.4.021017,Xiao,Poli}. It is claimed that there are interface state between two topologically distinct 1D systems. However the existence of interface state reflect no topological non-triviality.  Interface states can exist between any two periodic systems by choosing a suitable way to connect them. For example, interface state can exist between two trivial 1D systems as is shown in Fig.8. The two systems are the same bulk system with a relative displacement and the interface can be considered as a dislocation in a topologically trivial 1D system. So the existence of interface state need no topological protection and it only reflects the way we connect the two 1D systems.

\begin{figure}
  \includegraphics[width=8cm]{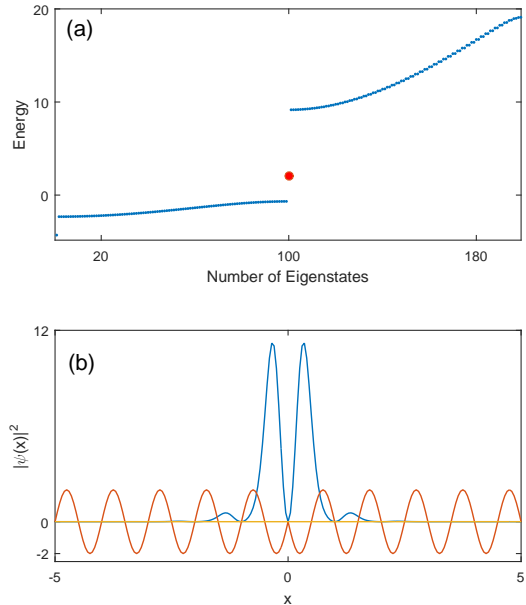}
    \caption{(a) Spectrum of two trivial periodic systems connected together. The potential $V(x)=V_0\sin(2\pi x)$ when $x>0$ and $V(x)=V_0\sin(2\pi (x-1/2))$ when $x<0$, where $V_0=-10$ and $m=\hbar=1$ in solving the energy eigenequation.(b)Density distribution of the interface state denoted by red dot in (a).}
    \label{Fig9}
\end{figure}

 As an example of the topologically triviality of interface state we consider the dielectric resonator chain in Ref. \onlinecite{Poli}. In this paper they claim there are topologically protected interface states between two topologically distinct configurations of SSH model with different winding properties. First we show these two configurations have no absolute meaning. If we consider the interface belong to the right system as is chosen in Ref. \onlinecite{Poli} then left system is in the $\alpha-$configuration and the right is in $\beta-$configuration(Fig.9(a)). However, if we consider the interface belong to the left system the conclusion is just the opposite(Fig.9(b)). These two choices are equally suitable to describe the chain. So there is no absolute distinction between $\alpha-$configuration and $\beta-$configuration and they are only the same bulk system with different choice of unit cells. A more symmetric and preferable way to describe the chain is depicted in Fig.9(c). In this case the interface site belong neither to the left not to the right system and it is just a dislocation of a single bulk system similar to that shown in Fig.8. So the existence of interface states reflect no topological distinction between the two systems.

 \begin{figure}
  \includegraphics[width=8cm]{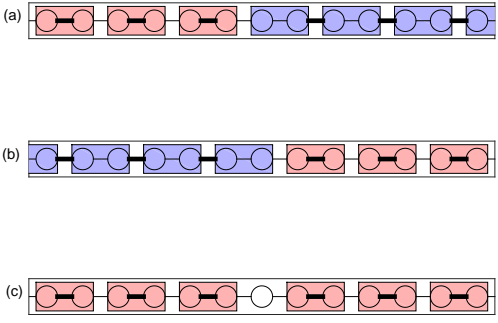}
    \caption{Three possible ways to describe the system considered in Ref. \onlinecite{Poli}. (a) is a reproduction of the schematic of SSH chain in Figure 1 (b) in Ref. \onlinecite{Poli}. Here we consider a SSH chain without absorption. So it is not necessary to distinguish the A and B sites. Both of them are depicted as same circles. Sites in the same box are considered to be in the same unit cell. System with red(blue) unit cell is considered as $\alpha-$configuration ($\beta-$configuration). Clearly, in this case the interface site should be considered as part of the right system. (b) and (c) The same SSH chain with different choices of unit cells. }
    \label{Fig8}
\end{figure}

 There are some works that study the topological property of 1D systems without considering the edge state property. As we shown above, there is no topological property at all if we do not consider the boundary property of the 1D system. For example, quantum walk has been used by several authors to discuss the topological properties in 1D systems. In Refs. \onlinecite{PhysRevLett.102.065703,PhysRevLett.115.040402} non-Hermitian systems are used to study topological transition. The winding number of the relative phase between two components of the Bloch functions are used as topological invariants to characterize to phases of the 1D model. The displacement $\langle\Delta m\rangle$ is proved to be quantized and is related to the winding property.

 However, the winding number depends on the choice of unit cell similar to the winding number for SSH model in Fig.2. Two equally possible choices of unit cell are depicted in Fig.10 (a) and (b). $\langle\Delta m\rangle$ and the winding numbers calculated with the two types unit cells are different. So $\langle\Delta m\rangle$  and the winding number only reflect the choice of unit cell and thus has no topological implication.

The unit-cell-dependence of $\langle\Delta m\rangle$ can be easily understood. In Refs. \onlinecite{PhysRevLett.102.065703,PhysRevLett.115.040402} sites in the same unit cell are labeled with same $m$. When a different unit cell is chosen the same site will be labelled with different $m$. So $\Delta m$ depends on the choice of unit cell. For example the intercell hopping with $\Delta m=1$ in Fig.10 (a) will be considered as intracell hopping with $\Delta m=0$ in Fig.10 (b). However the probability distribution for decay from different sites do not change with the different choice unit cell. So a different choice of unit cell will change $\langle\Delta m\rangle$.

\begin{figure}
  \includegraphics[width=8cm]{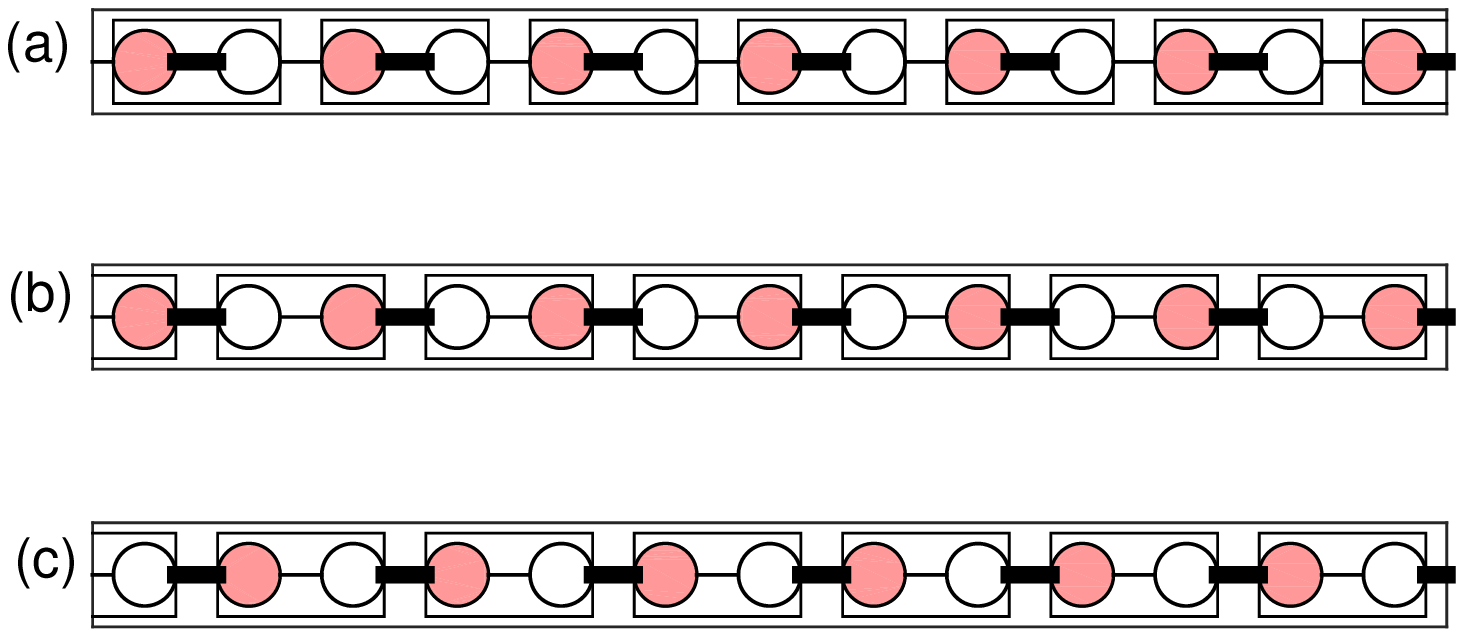}
    \caption{Three possible ways to describe the system in Refs. \onlinecite{PhysRevLett.102.065703,PhysRevLett.115.040402}. (a) is reproduction of the schematic of the biparticle lattice in Fig.1 (a) in Ref. \onlinecite{PhysRevLett.102.065703}. The thick and thin lines corresponds to the hoping amplitudes $v$ and $v'$.  Sites in the same box form a unit cell and should be labeled with same $m$ in Ref. \onlinecite{PhysRevLett.102.065703}. Filled circles denote the decaying sites. (b)The same system in (a) but with a different choice of unit cell.  (c) The same system in (a) viewed from opposite side. According to the theory in Refs. \onlinecite{PhysRevLett.102.065703,PhysRevLett.115.040402} (a) and (c) will be considered as two distinct phases.  }
    \label{Fig10}
\end{figure}

 In Refs. \onlinecite{PhysRevLett.102.065703,PhysRevLett.115.040402} they choose that the decaying site is on the left in one unit cell(though there is no reason why this is necessary). Even if we follow this instruction the so-called two topological distinct phases is still equivalent. If their conclusion is correct, two researchers  studying the same system will reach opposite conclusions on the topological property of the system just because they observe the same system from different sides as shown in Fig.10 (a) and (c). So the two so-called topological phases are just the same equivalent system viewed from different perspectives.

 In Ref. \onlinecite{PhysRevLett.113.046802} the Chiral-Symmetric AIII Class system is used to study the topological property of 1D system. The only difference between the model used Ref. \onlinecite{PhysRevLett.113.046802} and SSH model is the hopping amplitude between two orbits with same $n$ is pure imaginary. As shown above the phase of the hopping amplitude can be fixed arbitrarily when there are only nearest-neighbor hoppings. This model can be transformed to SSH model by a gauge transformation. So it is also a SSH model in disguise.  Similar to the SSH model, the $k$-space winding number $\nu$ of this system depends on the choice of the unit cell, so it is not a bulk topological invariant.

 A real-space expression of $\nu$ is used to study the effect of disorder on the so-called topological property of the system. After some derivation we find $\nu$ can be expressed as $\nu=\mathcal{T}\{(P_0-1)SX\}$, where $P_0$ is the projector to zero energy spectrum. There are many ambiguities in using this concept $\nu$. First, it depends on how the position operator $X$ is defined. The two orbits with same $n$ must be assumed to have the same position as in Ref. \onlinecite{PhysRevLett.113.046802} , otherwise $\nu$ will not be an integer even in the clean limit.

 Second, if $P_0=0$ it can be shown $\nu$ is identically zero in a system  with integer number of unit cells by noticing $\sigma_3$ is traceless and position operator is a constant at each site $n$. That is the two positions of the orbits in one unit cell cancel each other when we calculate the trance of $SX$. In a finite system, if there are even number of orbits in the system, there is no exact zero-energy state in the clean limit due to the coupling between the edge states at the two ends\cite{PhysRevLett.110.180403}. So, strictly speaking, $\nu$ should be $0$ and the disorder will not change this result.

 The simulation in Ref. \onlinecite{PhysRevLett.113.046802} get a finite $\nu$ in the clean limit perhaps because they consider the two edge states as being zero-energy. So edge states are not included in $P_+$ and $P_-$. In this case $\nu=\mathcal{T}\{P_0S_{+}X\}-\mathcal{T}\{P_0S_{-}X\}$. Edge state at the left(right) end of the system is eigenstate of $S$ of eigenvalue $1$($-1$). Thus, $\nu$ equals to the difference of average position of the two edge state divided by the length of the system, which is approximately $-1$. So it is a boundary property. When disorders present, especial when they are strong, the energy of the edge states will deviate from zero. So to calculate $\nu$ we must given a criterion how close the energy to zero should we consider it as zero-energy. Without such a criterion $\nu$ is meaningless. 

 In the clean limit, if we cut off the first and last orbits in the so-called topological phase. The two zero-energy edge states disappear and the system should be in the so-called trivial phase. However, we still get $\nu=-1$ if we use the original position operator $X$. In this case $P_0=0$ and $\nu=-\mathcal{T}\{SX\}$. $\nu$ does not vanish because there are single orbits at the two ends of the system and the position these two orbits remains when we calculate the trace of $SX$. Now we get an non-trivial $\nu$ when the system itself is trivial. To avoid this self-contradiction we must change the choice of unit cell and redefine the position operator to ensure two orbits in one unit cell have same position. After the redefinition of the position operator $\nu=0$. This again demonstrates $\nu$ is not a bulk property and it only depends on the choice of boundary.

 \section{Conclusions and outlook}

 In the present work we discuss the topological property of non-interacting 1D periodic systems. We find Zak phase is not a bulk topological invariant. In continuous model Zak phase only depends on the choice of origin of coordinate and there is no natural way to choose the origin of coordinate even when the system has inversion symmetry. We show Zak phase depends the choice of unit cell and gauge in the tight-binding model. The choice of unit cell and gauge of the bulk system can not be fixed by the chiral symmetry and inversion symmetry. So there is no topological non-triviality at all in most of the 1D systems if we do not look at the boundary of the system.

\begin{figure}
  \includegraphics[width=8cm]{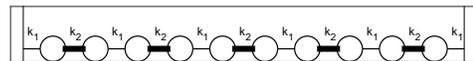}
    \caption{Schematic of coupled oscillators used to simulate the SSH model. All the oscillators have mass $m$. The springs are denoted by lines. $k_1$ and $k_2$ are coefficients of restitution of the springs. The springs at the boundary are attached to fixed walls. The system is assumed to be frictionless. When $k_1>k_2$ there are edge states in this system. }
    \label{Fig10}
\end{figure}

We show zero-energy edge states protected by chiral symmetry are only a boundary property, that is, they only reflect the choice of boundaries of the same bulk system.
Recently, there are some works trying to observe such edge states\cite{PhysRevLett.120.113901} of the SSH model. We find this kind of edge states can be most easily and accurately observed in coupled harmonic oscillators. For example the Hessian matrix of the system depicted in Fig.11 have the same form as the Hamiltonian of the SSH model except a term proportional to identity. This term only trivially breaks the chiral symmetry\cite{PhysRevB.84.195452}. In this system $\omega^2-\sqrt{(k_1+k_2)/m}$ corresponds to the energy of the SSH model, where $\omega$ is the eigen angular frequency of the system. The zero-energy edge states can be easily observed in this system. Though the boundary-dependent topological property can not be defined when there are odd number of oscillators, zero-energy edge state can be more easily observed in this case because there is no coupling between edge states at different ends. Theoretically, the edge state should have strictly zero-energy because there must be a state with strictly zero-energy in a chiral symmetric system when there are odd number of freedoms in it.

\bibliography{prb}
\end{document}